\documentclass[journal=jpccck,layout=twocolumn,manuscript=article]{achemso}

\usepackage{chemformula} 

\usepackage{bm}
\usepackage{braket}
\usepackage{amsmath}
\usepackage{amsmath}
\usepackage{amssymb}

\newcommand{\s}[1]{{\mathbf S}_{#1}}

\newcommand{\ez}{B}
\newcommand{\kavli}{\affiliation{Kavli
   Institute of Nanoscience, Delft University of Technology, 2628 CJ Delft, The Netherlands}}
\newcommand{\mff}{\affiliation{Department
  of Condensed Matter Physics, Faculty of Mathematics and Physics,
 Charles University, Ke Karlovu 5, 121 16 Praha 2, Czech Republic}}

\author{Peter Zalom}
\mff
\altaffiliation{Institute of Physics, The Czech Academy of Sciences, Na Slovance 2, CZ-18221 Praha 8, Czech Republic}
\author{Joeri de Bruijckere}
\kavli
\author{ Rocco Gaudenzi}
\kavli
\alsoaffiliation{Max Planck Institute for the History of Science, Boltzmannstrasse 22, 14195 Berlin, Germany}
\author{Herre S.~J. van der Zant}
\kavli
\author{Tom\'a\v{s} Novotn\'y}
\mff
\author{Richard Koryt\'ar}
\mff
\email{korytar@karlov.mff.cuni.cz}

\title{Magnetically-Tuned Kondo
Effect in a Molecular Double Quantum Dot: Role of the Anisotropic
Exchange}

\begin{document}


\begin{abstract} 
We investigate theoretically and experimentally 
the singlet-triplet Kondo effect induced by a magnetic
field in a molecular junction. Temperature dependent conductance, $G(T)$, is
calculated by the numerical renormalization group, showing a strong
imprint of the relevant low energy scales, such as the Kondo temperature,
exchange and singlet-triplet splitting. We demonstrate the stability of
the singlet-triplet Kondo effect against weak spin anisotropy, modeled
by an anisotropic exchange. Moderate spin anisotropy manifests itself
by lowering the Kondo plateaus, causing the $G(T)$ to deviate from 
a standard temperature dependence, expected for a spin-half Kondo effect.
We propose this scenario as an explanation for anomalous $G(T)$,
measured in an organic diradical molecule coupled to gold contacts. We uncover
certain new aspects of the singlet-triplet Kondo effect, such as
coexistence of spin-polarization on the molecule with Kondo screening
and non-perturbative parametric dependence of an effective magnetic
field induced by the leads.
\end{abstract}

\section{Introduction}

Electronic transport through single molecules with open shells 
allows the investigation of many fascinating phenomena which are rooted in 
the physics of the Coulomb blockade. A prominent example is the observation
of an underscreened Kondo
effect on a single entity, the Au+C$_{60}$ junction \cite{Roch2009}.
Other examples are the SU(4) Kondo effect \cite{Mugarza2012, Minamitani2012}
or a quantum phase transition driven by the gate voltage \cite{Roch2008}.
The reproducible and sharply defined chemical structure of molecules could
unveil new aspects of the Coulomb blockade physics, such as 
many-body quantum interference \cite{Mitchell2017}.

Molecules with two open-shell orbitals share certain features
with the so-called 
\emph{double quantum dots} (DQDs) and can be theoretically
 modeled by an
Anderson or Kondo model with two ``impurity'' spins.
These models exhibit a rich phenomenology, \emph{e.g.},
Refs.~\citenum{Alexander1964,Jayaprakash1981,Vojta2002,Hofstetter2001,
Varma2002,Langwald2019}.
Here we focus on a specific case when the two spins 
couple antiferromagnetically and are subjected to an external magnetic field.
The low-energy spectrum of such an
isolated molecule can be approximately captured by the Hamiltonian
\begin{equation}
\hat H_\mathcal M  = I \s{1}\cdot \s{2}
+ g\mu_\mathrm B\mathbf B\cdot
 \left (\s{1} + \s{2} \right),
\label{eq:hm}
\end{equation}
expressed through the spin operators $\s{1}$ and $\s{2}$.
The first term in Eq.~(\ref{eq:hm}) describes the antiferromagnetic interaction ($I>0$)
and the second term is the Zeeman term, corresponding to a homogeneous
magnetic field $\mathbf B$.
We show the magnetic field dependence
of the molecular spectrum in Scheme~\ref{s1}c, adopting the units $g\mu_\mathrm B=1$.
The ground-state has an accidental two-fold degeneracy if $|\mathbf B| = I$. 
The resulting effective two-level system, when coupled to leads,
exhibits the Kondo effect, as predicted in 
Refs.~\citenum{Pustilnik2000,Pustilnik2001}.
 
Recently, the DQD has gained renewed attention, because it can
host topologically-protected Weyl points. The Weyl points are particular
ground-state degeneracies which have incarnations in
diverse physical contexts, such as
molecular {conical intersections} \cite{Yarkony1996},
semi-metal band-structures \cite{Armitage2018}, or
 quantum field theory \cite{Volovik2003}.
In the DQD model, the Weyl points emerge when spin-orbit
effects are considered. Spin-orbit interaction effectively leads
to the addition of 
spin anisotropies in Eq.~(\ref{eq:hm}), namely, anisotropic
exchange interaction and anisotropic (and dot-dependent) $g$-tensor
\cite{Yosida,Herzog2010}.
As long as the anisotropies are
weak, a ground-state degeneracy can be found for
at least two magnetic fields $\pm\mathbf B_0$ related by
time-reversal.
These Weyl points were recently reported in InAs DQD\cite{Scherubl2018}
by means of a transport spectroscopy. When the magnetic field
is tuned to the degeneracy, \citeauthor{Scherubl2018}
observe a Kondo resonance.

Motivated by the significance of such magnetic-field induced level crossings,
we revisit the transport properties of the DQD 
near a degeneracy point. We present a combined experimental
and theoretical effort. In the experimental part of this work,
we show conductance measurements of a molecular junction:
a 2,4,6-hexakis-(pentachlorophenyl)mesitylene diradical bound
to gold contacts. This molecule (see Scheme~\ref{s1}a)
represents a prototypical
molecular DQD, where the two spins sit on the radical sites.
When the magnetic field is tuned to a
ground-state degeneracy, a Kondo-like zero-bias anomaly (ZBA) is observed.
Intriguingly, the temperature dependence of the ZBA does not follow a standard,
universal behavior of a Kondo impurity.

Thus motivated, we perform a comprehensive theoretical analysis of the
conductance of the DQD model in the vicinity of the 
magnetic-field induced ground-state degeneracy.
Our results include the effects of weak spin anisotropy.
We calculate the conductance by
the numerical renormalization group (NRG) technique.
Our results complement
earlier perturbative studies of the anisotropy effects in the
DQD \cite{Herzog2010,Stevanato2012}, because NRG
allows us to address quantum spin-fluctuations,
which eventually lead to the singlet-triplet Kondo effect. 
We offer a plausible and robust explanation of the anomalous temperature
dependence observed in our experiment. Moreover, we reveal
and analyze certain new aspects of the DQD, such as
coexistence of spin polarization and Kondo screening
at the degeneracy point and effective magnetic field induced 
by the leads.

\section{Methods}
\begin{scheme*}
	\includegraphics[width=\textwidth]{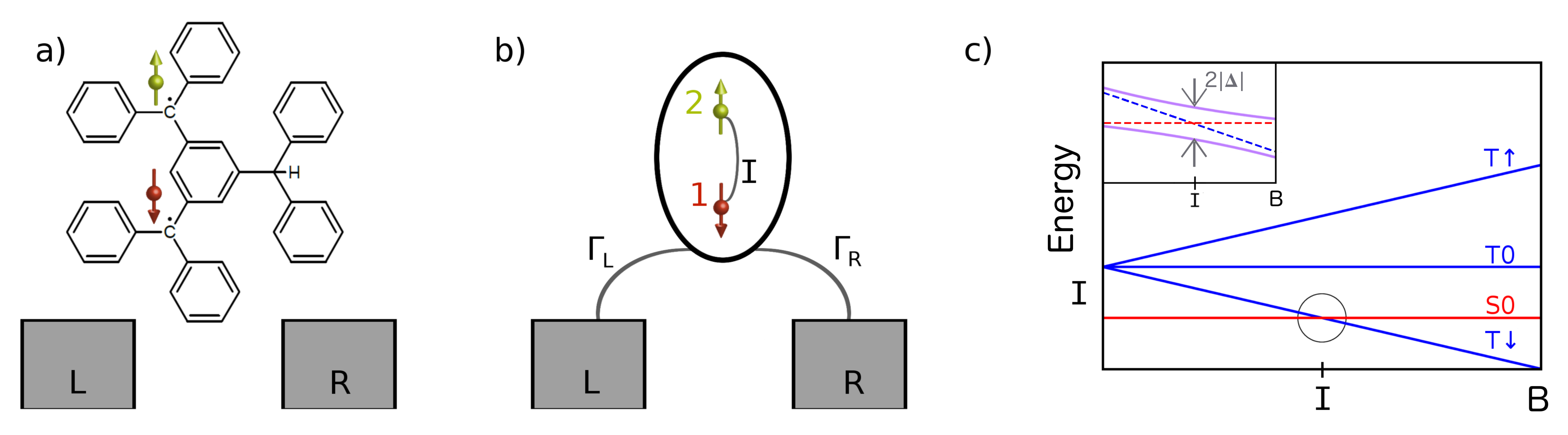}
	\caption{
(a) Illustration of a diradical molecule
[2,4,6-hexakis-(pentachlorophenyl)mesitylene] coupled to a pair
of leads.
(b) Schematic representation of the Hamiltonian: dot 1 couples to the leads, each
lead introduces single-particle level broadening $\Gamma_\mathrm L$,
$\Gamma_\mathrm R$. The dot 2 couples to dot 1 with
an anti-ferromagnetic exchange.
(c) Magnetic field dependence of the four lowest
 energy levels (eqs \ref{states}) of the double quantum
dot. The accidental
singlet-triplet degeneracy occurs when $B=I$. Inset: zoom
of the crossing. When Dzyaloshinskii-Moryia interaction, Eq.~\eqref{eq:ax},
is considered, the original crossing (dashed lines) turns into
an avoided crossing (purple lines) with the splitting  $2|\Delta|$.}
\label{s1}
\end{scheme*}

\subsection{Theoretical Methods}
\subsubsection{Double Quantum Dot in a Magnetic Field}
We introduce a model of a molecule coupled to two
leads in which
the singlet-triplet degeneracy can be achieved
by an external magnetic field. 
We shall assume that
the low-energy excitations are spin-excitations
due to exchange-coupled spins residing in two 
orbitals (two ``quantum dots'').
Such a model can
apply to organometallic complexes with
two open-shell transition-metal centers, organic diradicals
(see an example given in Scheme~\ref{s1}) and other open-shell
molecules with even number of electrons (for instance, Ref.\cite{Roch2008}).

Coupling the molecule to a pair of leads induces Kondo correlations, which
generally involve two screening channels \cite{Pustilnik2001,Mitchell2017}.
The two screening channels can be characterized by two characteristic
temperature scales (Kondo temperatures), $T_1$ and $T_2$ \cite{Pustilnik2001}.
In a general molecular-electronic setup,
the coupling of the first and second dots to the leads is highly asymmetric,
implying an exponential separation of $T_1$ and $T_2$. Unless $T_1\approx T_2$,
the stronger coupled
channel wins and the low-temperature
behavior is equivalent to a fully-screened Kondo impurity. 
Thus, in a major portion of the parameter space the physics is of a single-channel
type.
We shall consider in this work single-channel effects only, and for
this reason we can make simplifying assumptions on the details
of the dot-lead couplings.
Specifically, we will consider that only the first dot
couples to the two leads, while the coupling of the second dot to the
leads vanishes. We may thus disregard charge-fluctuations
on the second dot and consider only its spin degree of freedom,
represented by operator $\mathbf S_2$.

The Hamiltonian of the DQD can be written as $H_\text{DQD} = H_1 +
H_\text{ex} + H_\text{Z}$, where the individual terms read
\begin{subequations}
\label{eq:h_dqd}
\begin{align}
H_1
&=
\sum_{\sigma}
\varepsilon_d 
d^{\dagger}_{\sigma}
d^{\vphantom{\dagger}}_{\sigma},
\, \, + \, \,
U n_{d,\uparrow}, n_{d,\downarrow}
\\
H_{\mathrm{ex}}
&=
I \, \mathbf{S}_1 \cdot \mathbf{S}_2
\\
H_{\mathrm{Z}} &=  \ez \left( \hat S_{z,1} + \hat S_{z,2} \right).
\end{align}
\end{subequations}
The term $H_1$ represents the first dot as an Anderson impurity.
The operator $d^{\dagger}_{\sigma}$ creates an electron of spin $\sigma \in \uparrow,
\downarrow$, $n_{d,\sigma} = d^{\dagger}_{\sigma}
d^{\vphantom{\dagger}}_{\sigma}$ is the number operator,
$\varepsilon_d$ is the onsite energy and $U$ is
the charging energy.
The term $H_\text{ex}$ represents the antiferromagnetic
 exchange interaction ($I>0$) between
both dots. The operator $\mathbf{S}_i$ is the spin operator of
the respective dots. The operator $\mathbf{S}_1$ can be expressed
in terms of Pauli matrices $(\sigma_x,\sigma_y,\sigma_z)=\bm\sigma$
as $\mathbf{S}_1 = \frac 12
\sum_{\sigma''\sigma'} d^{\dagger}_{\sigma'} \bm\sigma_{\sigma'\sigma''}
d^{\vphantom{\dagger}}_{\sigma''}$.
Finally, the term $H_\mathrm Z$ represents
the homogeneous magnetic field in the $z$-direction;
the field strength $B$ is represented in units of energy 
(\emph{i.e.} $g\mu_\mathrm B=1$).

We investigate the properties of the DQD in the Coulomb
blockade regime, \emph{i.e.} when the occupancy of the first dot
is approximately one. Hence, the following
hierarchy of energy scales is assumed: $U, |\epsilon_d| \gg |B|,I$
and $\varepsilon_d < 0< U$. Consequently, the lowest-lying
eigenstates of $H_\text{DQD}$ are triplet and singlet and their energies read
\begin{subequations}\label{states}
\begin{align}
\ket{T\!\uparrow}  &=\ket{\uparrow \uparrow}, & E_{T\uparrow} &=\varepsilon_d + B + \frac{I}{4}\\
\ket{T0} &= \frac{1}{\sqrt{2}} \left(\ket{\uparrow \downarrow} + \ket{\downarrow \uparrow}\right), &  E_{T0} &= \varepsilon_d + \frac{I}{4}\\
\ket{T\!\downarrow} &= \ket{\downarrow \downarrow}, &  E_{T\!\downarrow} &=\varepsilon_d - B + \frac{I}{4}\\
\ket{S0} &= \frac{1}{\sqrt{2}}\left( \ket{\uparrow \downarrow} - \ket{\downarrow \uparrow}\right), &  E_{S0} &= \varepsilon_d - \frac{3I}{4}
\end{align}
\end{subequations}
where in the symbol $\ket{\uparrow\downarrow}$ the first
(second) arrow represents the spin projection of the
first (second) dot, respectively. The spectrum of $H_\text{DQD}$
is shown in Scheme~\ref{s1}c, where we can recognize the
ground-state degeneracy point at $B=I$.

\subsubsection{Anisotropic Exchange}
The Hamiltonian $H_\text{DQD}$ introduced here enjoys rotational
invariance in the spin space. This is an approximation,
because the spin is not a good quantum number due the to
spin-orbit interaction (SOI). We shall assume that the latter is
associated with the smallest energy scale (compared to $U$ and $I$),
which holds true for, e.g., organic molecules. The presence
of weak SOI can be accounted for by anisotropies in the exchange ($H_\text{ex}$)
and Zeeman ($H_\text{Z}$) terms \cite{Yosida}. Since we
consider effects related to the singlet-triplet crossing,
the main effect of the anisotropies will be to split the
degeneracy of the $\ket{S0}$ and $\ket{T\!\downarrow}$.
As a function of $B$, the crossing becomes avoided,
as shown in Scheme~\ref{s1}c.

Without loss of generality, we can consider a specific form
of the anisotropy, the Dzyaloshinskii-Moriya interaction
\begin{align}
\label{eq:ax}
H_\mathrm{A}
&= -2 \sqrt{2} \, \Delta \,
\left( \, \,
\hat{S}_1^x \, \hat{S}_2^z \, - \, \hat{S}_1^z \, \hat{S}_2^x \, \,
\right),
\end{align}
where $2|\Delta|$ yields the singlet-triplet gap at $B=I$.
The above interaction exhibits a special direction, the $y$-axis, 
which is commonly referenced to as a Dzyaloshinskii-Moriya vector.
We note that the level crossing induced by $H_\mathrm A$
 can be restored by rotating the magnetic field to the $y$-axis
\cite{Herzog2010}.

\subsubsection{Coupling to the Leads and Conductance}
The complete Hamiltonian of the molecule coupled to 
(left and right) leads
 consists of three terms $H = H_\text{DQD} + H_\mathrm l
+ H_\mathrm t$, where the subscript labels denote the double
quantum dot, leads and tunneling. The lead Hamiltonian has
a standard form
\begin{equation}
H_\mathrm l = \sum_{x,{k}\sigma} \varepsilon_{x,{k}}
c^{\dagger}_{x,k\sigma}
c^{\vphantom{\dagger}}_{x,k\sigma},
\end{equation}
where $c^{\dagger}_{x,k\sigma}$ is a canonical creation operator
and $\varepsilon_{x,{k}}$ are single-particle energies.
The indices denote spin $\sigma$, lead $x=\mathrm L,\mathrm R$
(for left and right) and
the remaining quantum numbers (e.g. bands and wave-numbers)
are encapsulated in $k$. The coupling
between the leads and the DQD is given by the tunneling
Hamiltonian $H_{\mathrm{t}}$ of the form 
\begin{equation}
H_{\mathrm{t}} =
\sum_{x,{k}\sigma} V_{x,{k}}
c^{\dagger}_{x,{k}\sigma} 
d^{\vphantom{\dagger}}_{\sigma}
+ \text{h. c.},
\label{eq:tun}
\end{equation}
where $V_{x,{k}}$ is the hybridization matrix element 
and h.c. stands for Hermitian conjugate.
Each lead gives rise to a single-particle hybridization
function defined by $\Gamma_x(\varepsilon) = 
\pi \sum_{\mathbf{k}} |V_{x,\mathbf{k}}|^2
\delta(\varepsilon-\varepsilon_{x,\mathbf{k}})$.
We employed hybridization functions that are
constant within a bandwidth $2D$, \emph{i.e.}:
$\Gamma_x(\varepsilon) = \Gamma_x \theta(D-|\varepsilon|)$.

Near the degeneracy point, the Hamiltonian $H$
[Eqs.(\ref{eq:h_dqd}, \ref{eq:ax}-\ref{eq:tun})] describes
a single-channel Kondo problem. As stated earlier, a non-vanishing
coupling of the dot 2 to the leads would imply a two-channel
problem, however, the latter is not commonly expressed in molecular
junctions, as the dominant screening channel takes over, so the
problem is effectively single channel.
A further consequence of having
both dots coupled to conduction electrons is that an
 effective exchange coupling 
$I_\text{RKKY}\, \mathbf{S}_1 \cdot \mathbf{S}_2$
of the Rudermann-Kittel-Kasuya-type emerges\cite{Vojta2002}.
Hence, the effect of the coupling of dot 2 to leads can be
seen as a renormalization of the exchange $I$.

For the (linear) conductance the following relationship
holds
\begin{multline}
G(T) =
\frac{2e^2} h
  \frac{4\Gamma_\mathrm L\Gamma_\mathrm R}
{(\Gamma_\mathrm L + \Gamma_\mathrm R)^{2}} \\
\times\int_{-\infty}^\infty\mathrm d\omega\ \pi
\Gamma A_1(\omega) \big[-n'_\mathrm F(\omega)\big]
\label{eq:gt}
\end{multline}
where $\Gamma \equiv \Gamma_\mathrm L +\Gamma_\mathrm R$, the 
derivative of the Fermi-Dirac
distribution is denoted by $n'_\mathrm F$, and $A_1(\omega)$ is the
spectral function of the first dot. The only
effect of the asymmetry of the couplings to both leads is
to modify the prefactor of the integral in Eq.~\eqref{eq:gt}.
This motivates us to introduce the  conductance unit
\begin{equation}\label{G0}
G_0 \equiv \frac{2e^2} h
  \frac{4\Gamma_\mathrm L\Gamma_\mathrm R}
{(\Gamma_\mathrm L + \Gamma_\mathrm R)^{2}}.
\end{equation}

\subsubsection{Estimates of the Energetic Scales in the Molecular Problem}
The model that we introduced is based on considerable simplifications
of the electronic structure of the molecule coupled to leads.
The simplifications can be justified by the fact that the emergent low-temperature
Kondo physics is always governed by only few parameters (\emph{e.g.} $T_\mathrm K$,
$G_0$, $I$, $\Delta$). However, these parameters do not directly relate to 
the energy scales of the molecule in isolation due to interactions between the
molecule and contacts. We give our estimates in what follows.

The exchange coupling $I$ characteristic of isolated
organic diradicals can be (typically) 40 meV $> I >$ 0.4 meV \cite{Rajca1994}.
As we remarked, the hybridization of both ``dots'' with the leads
can slightly renormalize $I$.
The parameter $\Delta$ causes spin anisotropy, and can be estimated by
zero-field splittings. For the diradicals, the values $\approx 50 \mu$eV
have been reported, for instance in Ref.~\cite{Gallagher2016}.
The value of $U$ can be obtained from charging energy in the gas-phase, however,
the latter is considerably screened by the lead electrons. We estimate the value
of the order of $U\approx 100$ meV. The energy scale $|\varepsilon_\mathrm d|$
of an Anderson impurity is, in principle, the approximate ionization energy of the molecule
coupled to the leads. It is unfortunately not possible to estimate $\varepsilon_\mathrm d$
from gas-phase ionization levels because the alignment of the ionization level
with the Fermi energy of the metal contacts is affected by partial charge transfer.
Moreover, some molecular transport experiments operate with a gate voltage,
allowing the effective tuning of the value of $\varepsilon_\mathrm d$.
The single-particle energetic broadening $\Gamma$
is on the order of 5 meV\cite{gaudenzi2017redox} and it is sensitive to the binding geometry. 

\subsubsection{Numerical Renormalization Group Calculations}

For the numerical analysis of the present double-dot model, we have utilized
the open-source code NRG LJUBLJANA \cite{NRGljubljana,ZitkoPruschke-2009}. The spectral functions
have been obtained by the full density matrix algorithm based on the complete
Fock-space concept \cite{NRGljubljana, Weichselbaum2007}. The interleaved
method has been used to smoothen the resulting spectral functions
\cite{NRGljubljana, Weichselbaum2007} while the logarithmic discretization
parameter has been set to $\Lambda=2$.  All results
are obtained for $\varepsilon_d = -U/2$.

\subsection{{Experimental Methods}}
The molecule used here is a 2,4,6-hexakis-(pentachlorophenyl)mesitylene diradical
\cite{veciana1993stable} depicted in Scheme~\ref{s1}a.
The single-molecule device used for the transport measurements is similar to
the one used in Ref.~\citenum{gaudenzi2017redox}. By electromigration \cite{park99} and
self-breaking \cite{oneill07} of a gold nanowire, a nanometer-sized gap is
formed, in which molecules can be trapped to realize a single-molecule
junction. After electromigration, a dilute solution of the molecules of
interest is drop-cast on a chip with 24 electromigrated gold nanowires. After
pumping away the solution and cooling down the system in a dilution
refrigerator ($T \approx$ 40 mK), we typically find transport signatures of
single molecules in 2 to 5 junctions per chip. We measure the DC current $I$
through the single-molecule devices as a function of the applied bias voltage
$V$ over the junction, the voltage applied to a capacitively coupled gate
electrode $V_\text{g}$, the temperature $T$ (20~mK $<T<$ 4.2~K),
and the magnetic field $B$. 


\section{Results and discussion}

\subsection{Temperature Dependence of the Conductance}
\begin{figure}
\includegraphics[width=9cm]{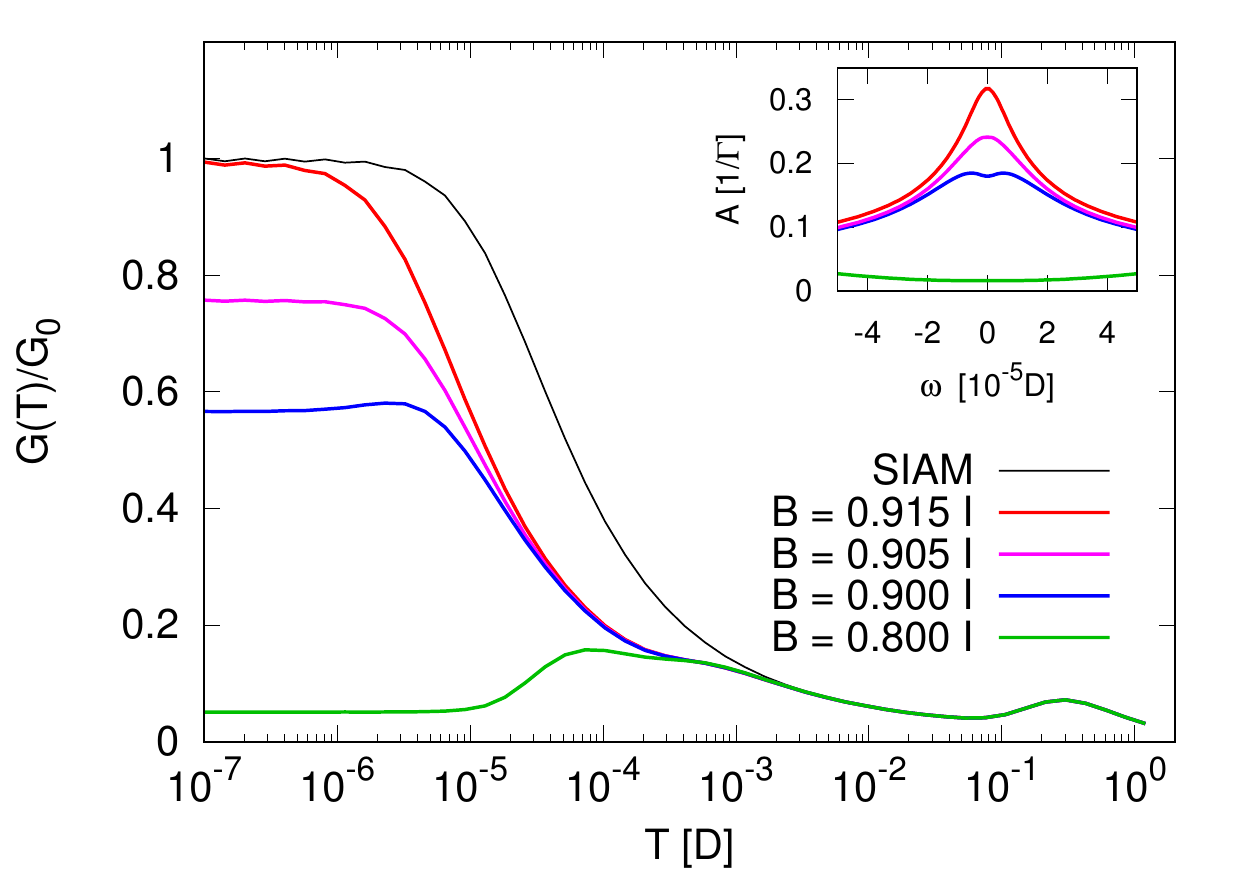}
\caption{NRG results on the double-dot model: temperature dependence
 of the conductance in units of $G_0$ (eq \ref{G0}) for the exchange
coupling $I = 10^{-3} D$, broadening
$\Gamma = 0.05D$ and various values of $\ez$. For comparison,
we also show the conductance of the single-impurity Anderson
model ($I=\ez=0$). The inset shows the corresponding zero-temperature spectral functions.
\label{g1}}
\end{figure}

For the sake of reference, we start by presenting the conductance
of the SIAM. We note that when $\ez = I =0$ in Eq.~\eqref{eq:h_dqd},
the transport properties of the DQD are equivalent to the SIAM,
because the second dot is decoupled.
We choose the parameters  $U=D$ (used throughout the whole paper) and $\Gamma=0.05D$, which correspond
to the Coulomb blockade regime.
In Figure~\ref{g1} the black curve represents the temperature-dependent
conductance $G(T)$, which exhibits a familiar low-temperature plateau
due to the Kondo effect with $T_\mathrm K(\text{SIAM})
 \approx 5\cdot 10^{-5} D$ (estimated from $G(T_{\mathrm K})=G_{0}/2$).
For intermediate temperatures the conductance is suppressed, until
$T$ reaches the temperature scale of the charge excitations $U/2$.

In the next step we couple the second dot: we choose $I = 10^{-3}D$
so that the lowest-lying states  of the isolated DQD are
$\ket{S0}$ and $\ket{T\sigma'}$, \emph{i.e.} singlet and triplet states.
Figure~\ref{g1} shows the conductances for different values of the
Zeeman energy $\ez$, chosen so that $\ez \sim I$. The high-$T$
part of $G(T)$ is almost identical to SIAM, the differences
show up at low temperatures, when $T \lesssim I$. For $\ez = 0.8I$ 
the ground state of the isolated DQD
is $\ket{S0}$ and the lowest-lying excited state is $\ket{T\!\downarrow}$,
separated by an energy gap $I-\ez > T_\mathrm K(\text{SIAM})$. 
The Kondo plateau is thus suppressed for this value of $\ez$.
 The bump for $10^{-5} < T
< 10^{-2}$ corresponds to energy scales of spin excitations. Indeed,
within the independent-particle picture,
the elevated conductance can be traced to two effects: the thermal
population of the $\ket{T\sigma'}$ states and the thermal broadening
of the Fermi distribution of conduction electrons (see, e.g.,
Ref.~\citenum{Korytar2012}).

When increasing the value of $\ez$ toward $I$, the singlet-triplet
degeneracy point is approached and the low-temperature plateau emerges.
When $\ez = 0.9 I$, the spectral function shows a split-peak
(inset of Figure~\ref{g1}), in qualitative
agreement with Ref.~\citenum{Chung2007}.
For $\ez = 0.915 I$, $G(0)$ reaches the unitary limit (red curve 
in Figure~\ref{g1}).
When the temperature dependences are compared to the SIAM,
 we observe two significant differences: First,
the Kondo temperature of the DQD is suppressed by a 
factor $\approx 10^{-1}$.
Second, the spin excitations give rise to the bump at $T\approx I$.

\subsection{Renormalization of the Resonant Magnetic Field}
\begin{figure}
	\includegraphics[width=1.05\columnwidth]{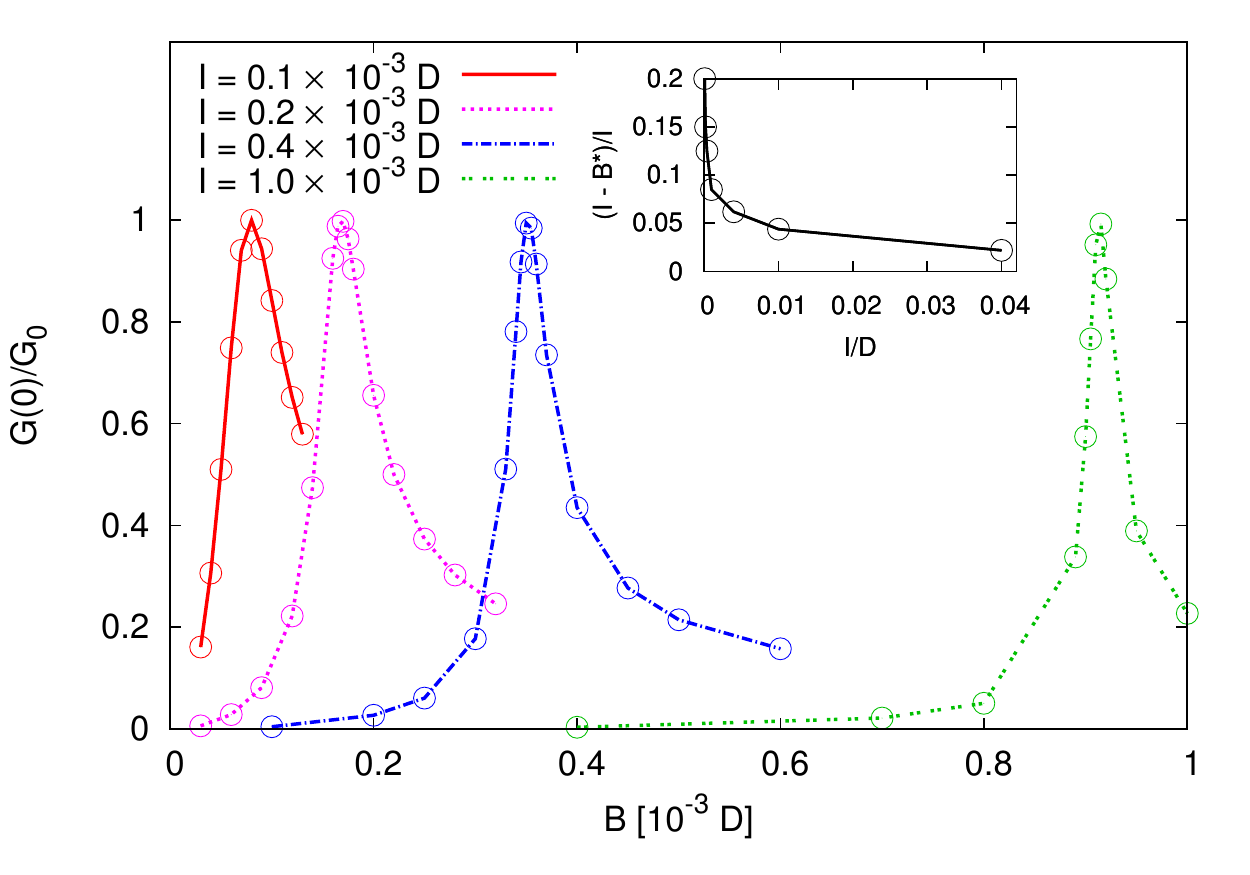}
\caption{Theoretical zero-temperature conductance $G(0)$ as
a function of the
external magnetic field $B$ for various values of $I$ ($\Gamma = 0.05D$
is fixed). The
locations of the maxima define the resonant field $\ez^*$. The difference
$I-\ez^*$ can be interpreted as an effective field generated by the leads.
Lines are only for visual guidance. The inset shows the dependence of the
normalized effective  magnetic field $(I-\ez^*)/I$ on $I$.  \label{g2}
}
\end{figure}

The above observations are consistent with the Kondo effect, which
is induced by magnetically tuning the DQD to a degeneracy point,
as predicted by \citeauthor{Pustilnik2000} in Ref.~\citenum{Pustilnik2000}.
The unitary conductance is, however, not observed at $\ez = I$,
as we demonstrate in Figure~\ref{g2}, where we show the zero-temperature conductance as a
function of $\ez$. We denote the location of the maxima of the conductance
as $\ez^*$ and observe that the latter are consistently
shifted toward lower values, below the ``bare'' degeneracy
condition $\ez = I$. The difference $I-\ez^*$ can be understood
as an effective magnetic field generated by the leads.
This effect was described as a shift of the degeneracy point
in Refs.~\citenum{Pustilnik2000,Pustilnik2001,Golovach2003,Chung2007, Herzog2010} but it
was not analyzed in more detail.\bibnote{In Ref.~\cite{Hock2013} the
effective field was analyzed in a hard-axis single spin impurity.}

\begin{figure}
	\includegraphics[width=1.05\columnwidth]{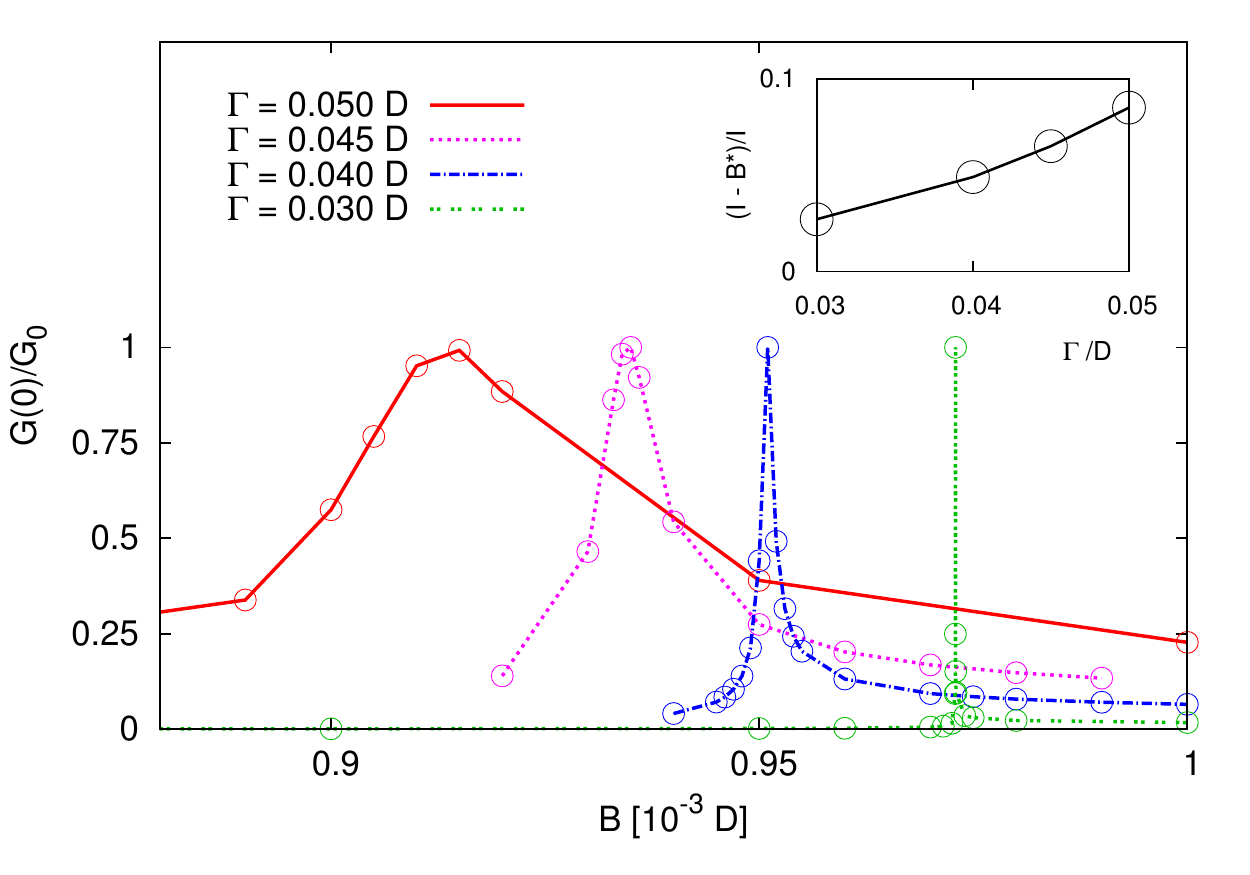}
	\caption{Theoretical zero-temperature conductance $G(0)$ as
a function of the
external magnetic field $B$ for various values of $\Gamma$ ($I=10^{-3}D$
is fixed).
Lines are only for visual guidance.  The inset shows the dependence of the
normalized effective  magnetic field $(I-\ez^*)/I$ on $\Gamma$.}
	\label{fig4}
\end{figure}

To inspect the effective magnetic field more closely, we fixed
$I$ and changed the hybridization strength $\Gamma$. The values 
of $G(0)$ plotted against the
external magnetic field $B$ are presented in Figure~\ref{fig4}. The
width of the resonant peak tends sharply to zero with decreasing $\Gamma$.
This observation can be rationalized by the concomitant (exponential)
decrease of $T_\mathrm K$, causing the Kondo resonance to be less
robust as the magnetic field departs from the degeneracy point.
The resonance field $\ez^*$ tends to the bare value $I$
as $\Gamma$ decreases, as shown in the inset of Figure\,\ref{fig4}.
The effective field can be fit by a power-law $I-B^*\propto
\Gamma^\alpha$ with $\alpha = 2.22 \pm 0.08$.

The presence of an effective magnetic field acting on the DQD is 
rooted in the fact that the two states $\ket{S0}$ and 
$\ket{T\!\downarrow}$ have a different orbital structure.
Thus, the leads renormalize the energies $E_{S0},E_{T\downarrow}$
in a different way. This can be contrasted with the SIAM,
where the spin-up and spin-down states are related by an inversion
of the spin quantization axis. The latter is a symmetry operation
of the leads. Consequently, no effective magnetic field can be 
generated in the SIAM. The intriguing parametric dependence
of $B^*$ points to a non-perturbative nature of the effective
magnetic field. 


\subsection{Spin-Polarization of the Kondo-Screened Dots}
\begin{figure}
	\includegraphics[width=1.05\columnwidth]{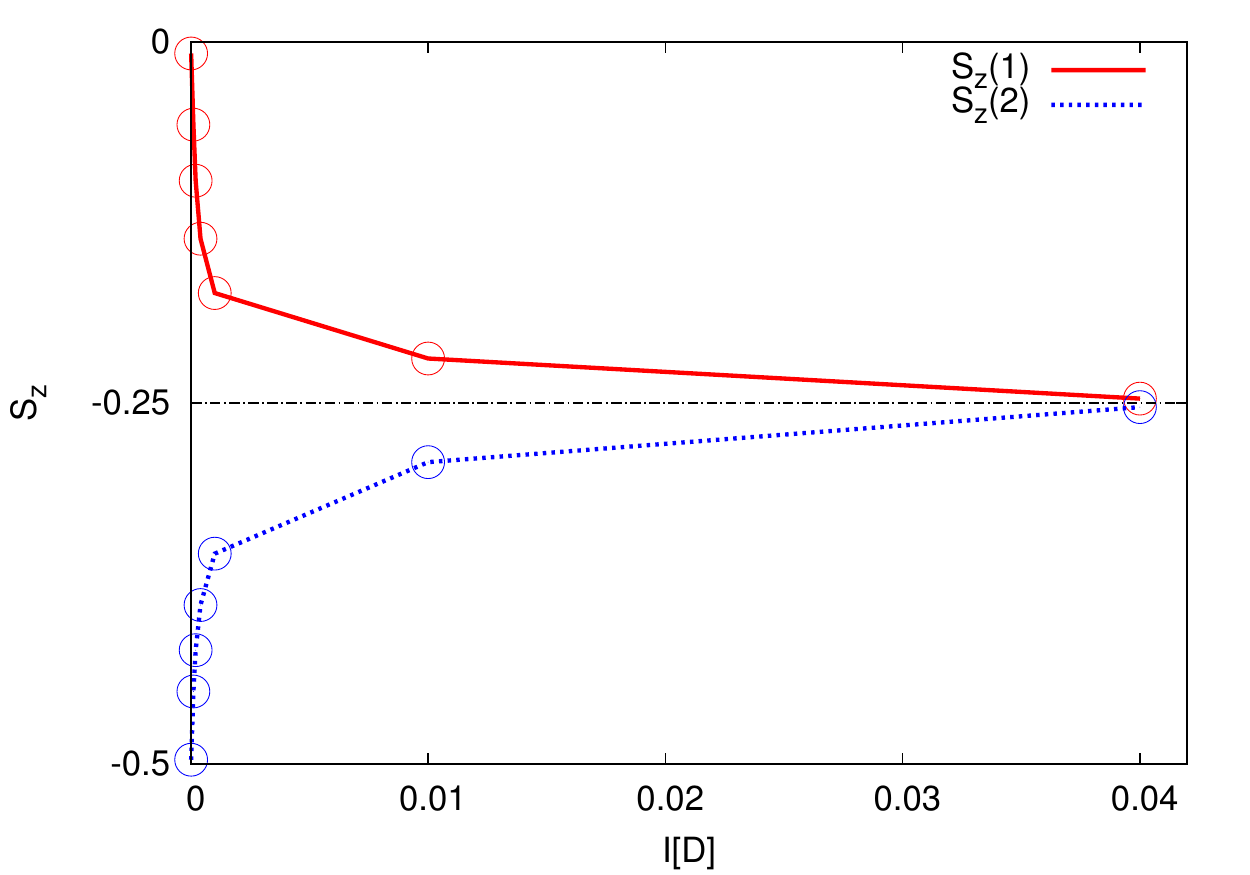}
	\caption{$z$-component of the spin
 of the primary (label $1$, red full curve)
and secondary (label $2$, blue dashed curve) dot as a function of $I$.
Lines are only for visual guidance.
}
	\label{g4}
\end{figure}

We explore another peculiar consequence of the broken
spin-inversion symmetry.
In Figure~\ref{g4} we plot the $z$-component of the spins of the two dots
as a function of $I$. We emphasize that the magnetic field is always tuned
to the resonance $B^*$, \emph{i.e.}, the conductance is unitary.
Surprisingly, despite the Kondo screening, the two dots exhibit
fractional spin polarization, which depends continuously on $I$.

We can understand the expectation values of spin in two simple
limits: $I \ll T_\mathrm K(\text{SIAM})$ and 
$I \gg T_\mathrm K(\text{SIAM})$.
For $I=0$ it is seen that $S_z(1) = 0$, as expected for the SIAM
with the first dot fully screened.
The second dot is decoupled and its spin aligns along the field.
When $0<I \ll T_\mathrm K(\text{SIAM})$, the second dot couples 
antiferromagnetically
to the local Fermi-liquid excitations of the first Kondo-screened dot.
It is known that when $\ez =0$, second-stage Kondo screening develops, with
a characteristic temperature $T_\mathrm K^{(2)}$,
below which the conductance is suppressed\cite{Hofstetter2001}.
In our case, the external magnetic field is always tuned
to achieve $G(0)= 1\cdot G_0$, so that the two-stage screening is avoided.

In the opposite limit of large $I$ we can see that 
both spins approach $-1/4$.
As long as charge-fluctuations can be neglected, the latter result
can be rationalized as follows: in the large-$I$ limit, the DQD can be
approximately described as a two-level system (TLS), the levels being
$\ket{S0}$ and $\ket{T\!\downarrow}$. The expectation values
of spin of either dot ($i=1,2$) in these states
are $\bra{S0}\hat S_z(i) \ket{S0}=0$ and
$\bra{T\!\downarrow} \hat S_z(i)\ket{T\!\downarrow}=-1/2$.
The interaction of the TLS
with the leads can be described by an anisotropic Kondo Hamiltonian,
as elaborated in Ref.~\citenum{Pustilnik2001}. When $B=B^*$ (and $T=0$)
there is the Kondo effect, so that
the two states have equal weights in the reduced density matrix.
It follows that the expectation value of spin on both dots
must be the equal average of $0$ and $-1/2$, \emph{i.e.}, $-1/4$.

In Figure~\ref{g4} we see that even when 
$I\gg T_\mathrm K \approx 10^{-4}D$, 
the deviation of $S_z(i)$ from $-1/4$ amounts to 10\% or more.
While the observation of Kondo plateaus (see Figure~\ref{g1}) can be
consistently accounted for by the anisotropic Kondo Hamiltonian,
other observables,
such as the spin, are not consistent with this model.
The spin polarization hints at sizable admixture of
 states $\ket{T0}$ and
$\ket{T\!\uparrow}$ in the many-body ground-state.
An effective low-energy model of the DQD should also
include the latter states in order to account for the
spin polarization.

\subsection{Effects of Anisotropic Exchange}
We have shown that despite the presence of the effective magnetic
field generated by the leads, the Kondo plateaus can be reached by tuning
the external magnetic field slightly away from the bare resonance
condition $B=I$. We show below that this does not hold true,
if the anisotropic exchange (AX) of the form in Eq.~\eqref{eq:ax} is introduced.

\begin{figure}
\includegraphics[width=1.05\columnwidth]{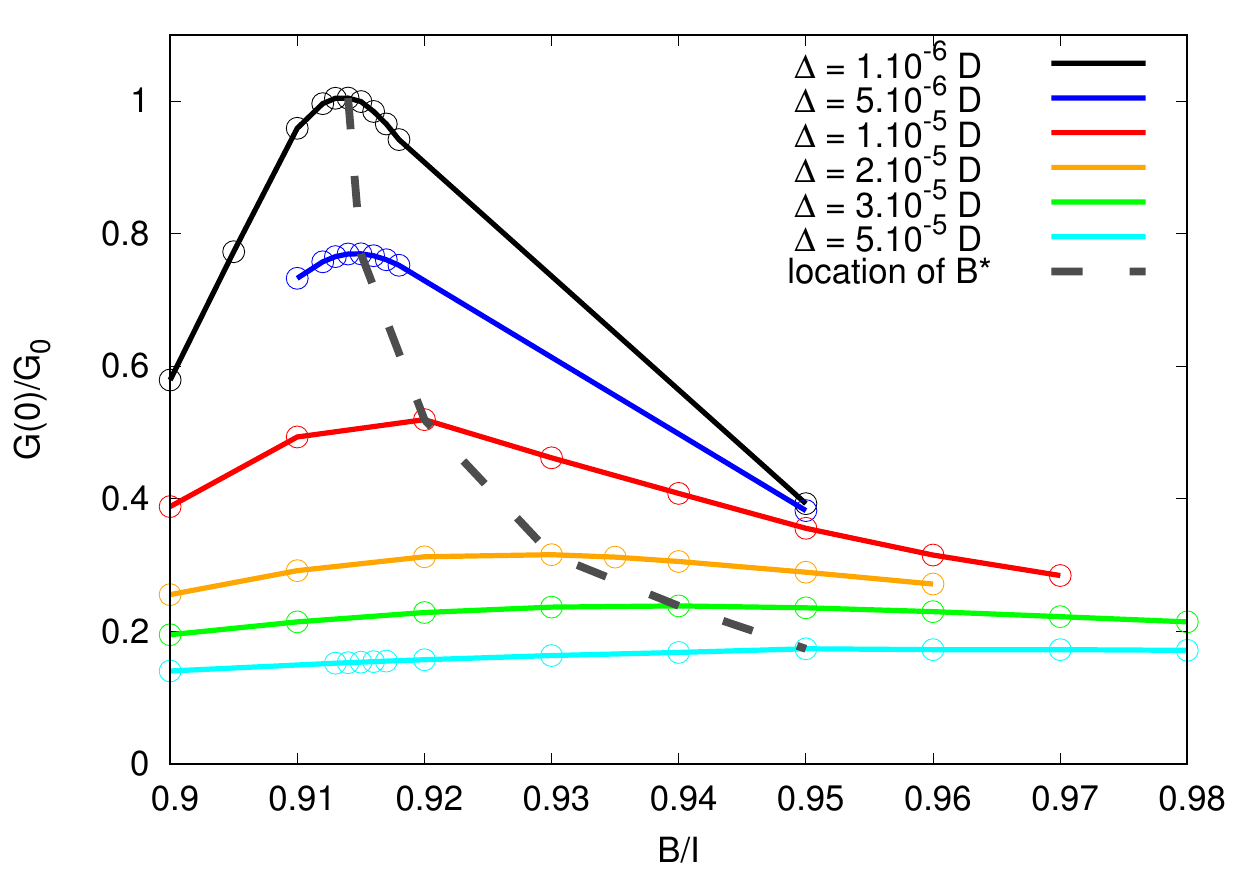}
\caption{NRG results on the double-dot model with anisotropic
exchange: Dependence of the zero-temperature conductance
 $G(0)$ on external magnetic field $B$
for various $\Delta$ ($I = 10^{-3}D$ and $\Gamma = 0.05D$). 
Lines are only for visual guidance.
\label{g5}}
\end{figure}

\begin{figure}
\includegraphics[width=1.05\columnwidth]{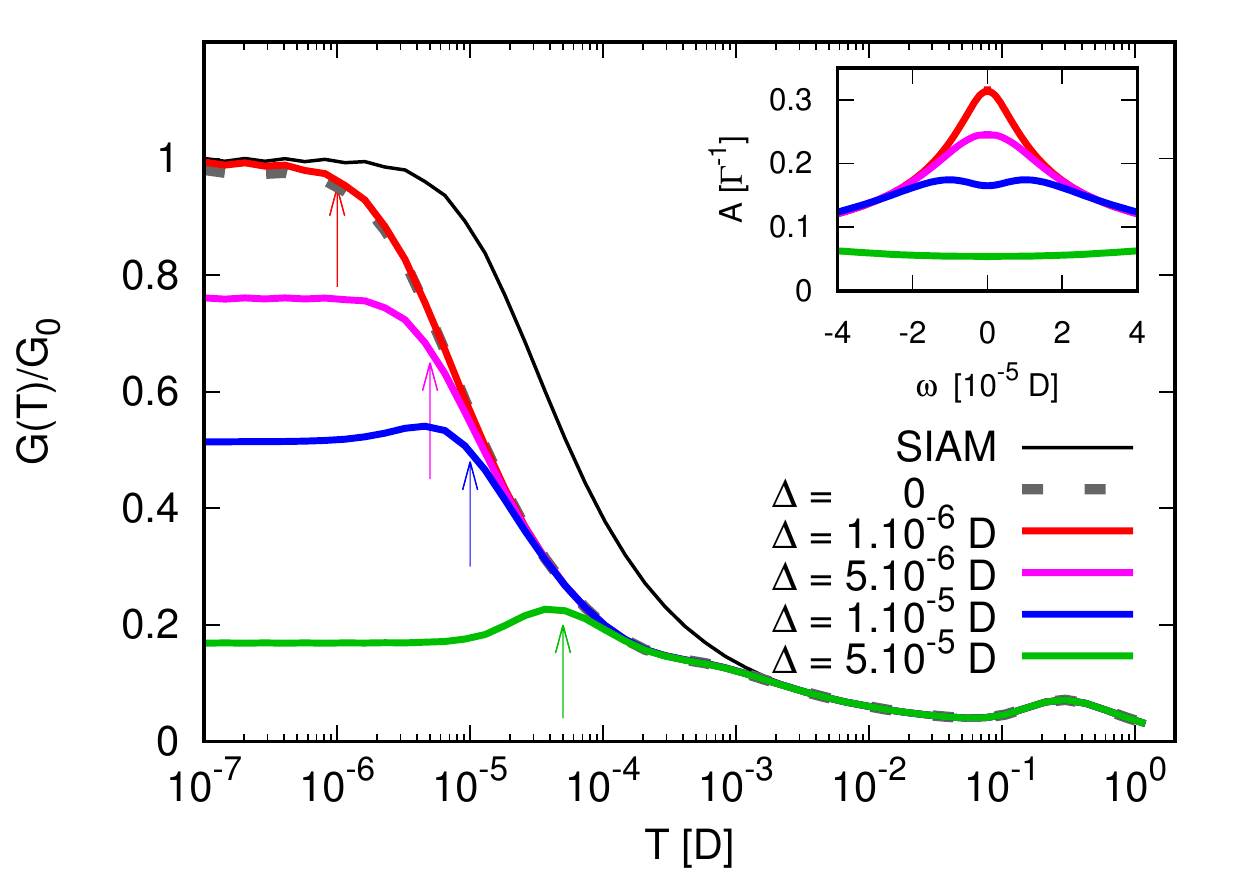}
\caption{NRG results on the double-dot model with anisotropic
exchange: 
Temperature dependence of the conductance for selected values
of $\Delta$ ($I = 10^{-3}D$ and $\Gamma = 0.05D$). The magnetic field has been
tuned to the corresponding value of resonant $B^*(\Delta)$ from Figure \ref{g5}.
The temperature equivalent to $\Delta$ is indicated by arrows.
The inset shows the spectral functions at zero-temperature.
}
\label{g6}
\end{figure}

Figure~\ref{g5} shows the dependence of the conductance at zero
temperature for a varying magnetic field. The effect of AX is
to lower the peak value of the conductance and shift its location.
The lowering of the peak value is caused by lowering of the Kondo
plateaus in the temperature dependence, as shown in Figure~\ref{g6}.

This behavior can be understood in a simpler physical
picture of the TLS
on the subspace spanned by the two lowest-energy states
$\ket{S0}$ and $\ket{T\!\downarrow}$. We shall denote the two states
by $\ket{\tilde\uparrow}$ and $\ket{\tilde\downarrow}$.
 The matrix elements
of $H_\mathrm A$ in the TLS are identical with the
 matrix elements of $\Delta {\sigma}_x$, where
 ${\sigma}_x$ is the Pauli
matrix. On the other hand,
 the energy gap between the two
states can be represented by $\frac 12\tilde B_z \sigma_z $.
The $\tilde B_z$
incorporates the bare splitting $I-B$, as well as
the effective field. In the TLS picture it is easy to see that
the AX destroys the Kondo effect and that the latter
can not be restored by tuning $B$, because
$\Delta \sigma_x$ causes an avoided level crossing.

We underline two important observations made from Figure~\ref{g6}:
First, the conductance plateaus are not destroyed by small
AX ($\Delta = 10^{-6}D<T_\mathrm K$). Second, the
Kondo peak diminishes and splits with increasing $\Delta$.
This behavior is reminiscent of the behavior of 
a standard spin-half Kondo impurity in a magnetic field.
There, a small
Zeeman splitting (much smaller than the Kondo temperature)
does not destroy the conductance plateau 
(it is  a ``marginal'' term) and leads to the peak splitting
\cite{Costi2000,Costi2001,Moore2000}. Concluding, we have shown that the
robustness of the Kondo plateaus is consistent with
the interpretation of the AX as a pseudo-magnetic field
acting on the TLS.

While in our calculation, we use only a specific form of the AX with
the Dzyaloshinskii-Moriya vector (DMV) aligned with the $y$-axis,
it can be seen that a general DMV translates into the TLS
as a pseudo-magnetic field, represented by a linear
combination of $\sigma_x$ and $\sigma_y$ matrices.
Only when the DMV is parallel with the $z$-axis (the direction of the external
magnetic field), the matrix elements of the AX in the TLS vanish.
The crossing is preserved in this special case, as pointed
out also in Ref.~\citenum{Herzog2010}.

\subsection{Anomalous Temperature Dependence of a Kondo Resonance
in a Diradical Molecule}

In the following, we present an experimental demonstration of the Kondo effect
at a singlet-triplet degeneracy, measured in the diradical single-molecule
junction described in the Experimental Methods. We show that the temperature
dependence of this Kondo effect strongly deviates from the standard spin-1/2
Kondo effect and we show that this deviation may stem from the anisotropic
exchange discussed in the previous section.

The diradical molecule used for these measurements consists of two unpaired
spins in its ground state. When embedded in a single-molecule junction, the
spins in the diradical molecule have a relatively weak exchange coupling $I
\sim 1$~meV. As a result, the energies of the spin singlet and one projection
of the spin triplet can become degenerate in an achievable magnetic field, 
as schematically
depicted in Scheme \ref{s1}c. This property opens up the possibility to experimentally
observe the singlet-triplet Kondo effect in the diradical molecule.

\begin{figure}
\includegraphics{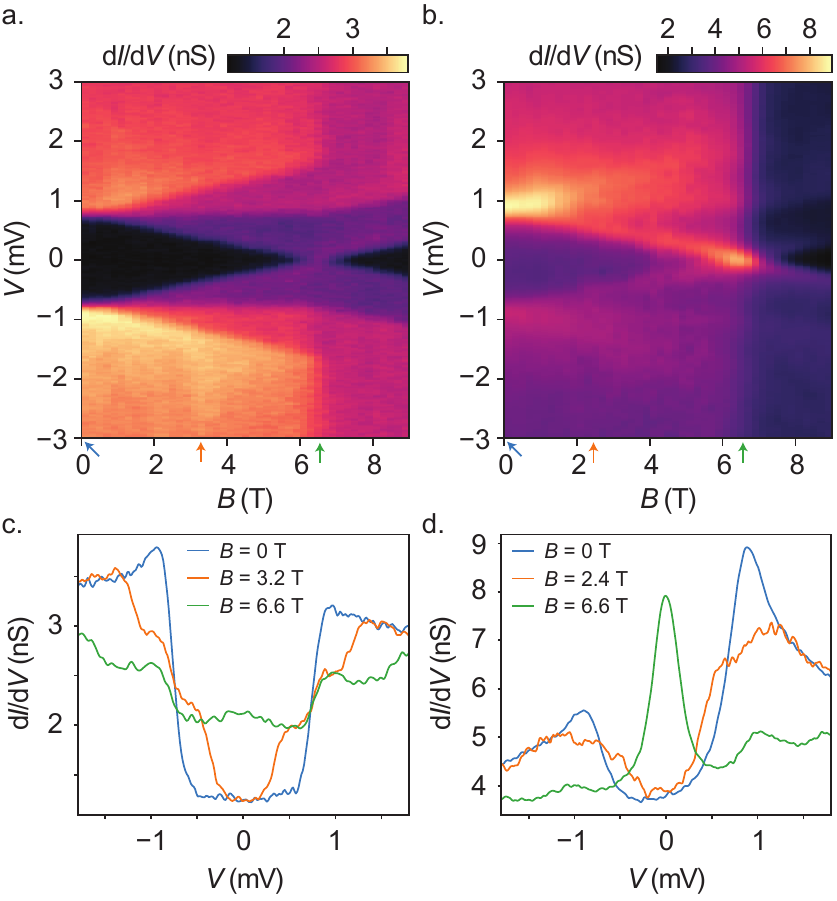}
\caption{(a,b): Experimental differential conductance (d$I$/d$V$) maps showing
the magnetic field evolution of the spin excitations between the singlet and
triplet states in the diradical single-molecule device, measured at (a)
$V_{\text{g}} = -1.7$~V and (b) $V_{\text{g}} = -2.8$~V. In part b, the
singlet-triplet Kondo resonance appears at $B\approx 6.6$~T. (c,d)
$dI$/d$V$ spectra at different magnetic fields, corresponding to vertical
line-cuts of the d$I$/d$V$ maps in (a) and (b), respectively. The magnetic
fields at which the spectra in (c) and (d) are recorded are indicated by the
colored arrows in (a) and (b), respectively.}
\label{g7}
\end{figure}

We probe the spin states of the single-molecule device by measuring the
differential conductance (d$I$/d$V$) as a function of $V$ and $B$. The results
of this experiment are shown in Figure~\ref{g7}, which contains two d$I$/d$V$ maps of
the same device, recorded at different gate voltages $V_{\text{g}} =
-1.7$~V (a) and $V_{\text{g}} = -2.8$~V (b). First, we focus on Figure \ref{g7}a, which
at $B=0$~T shows a stepwise increase in the d$I$/d$V$ at $V\approx \pm 0.7$~mV,
resulting from added transport channels involving excited states. The
excitation steps split as the magnetic field is increased and follow three
different slopes. This splitting is a clear manifestation of the Zeeman effect
in a spin system with a singlet ground state and a triplet excited state, with
an excitation energy equal to the exchange coupling $I\approx 0.7$~meV. At about
$B\approx 6.6$~T one projection of the triplet state becomes degenerate in energy with
the singlet state and at even higher magnetic fields this projection becomes
the new spin ground state. Only two spin excitations are observed after this
spin ground-state transition ($B\gtrsim6.6$~T), as expected from the spin selection
rules \cite{gaudenzi2017transport}.

By changing the gate voltage we were able to tune the single-molecule device closer to a
charge degeneracy point, which typically results in an increase of the overall
conductance and a stronger Kondo coupling. This behavior can be observed in the
d$I$/d$V$ map of Figure \ref{g7}b, which is recorded at $V_{\text{g}} = -2.8$~V. The
excitation steps seen in Figure~\ref{g7}a appear in Figure~\ref{g7}b as peaks rather than
steps. These peaks are fingerprints of higher-order transport processes,
which give rise to Kondo correlations \cite{ternes2015spin}. At the
singlet-triplet degeneracy ($B\approx 6.6$~T), a zero-bias resonance develops, which
can be attributed to the singlet-triplet Kondo correlations.

\begin{figure*}
\includegraphics[width=0.7\textwidth]{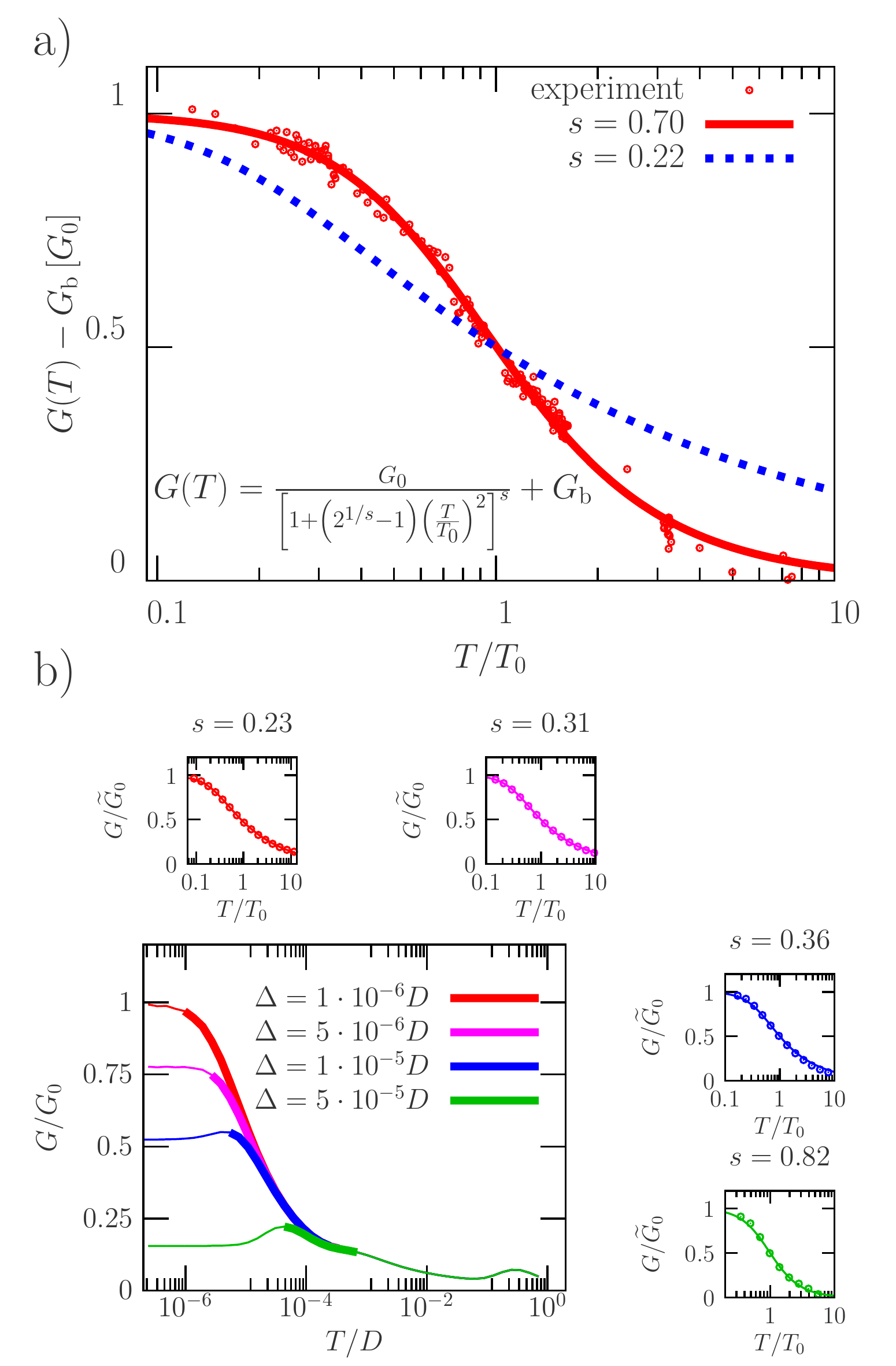}
\caption{\label{g8}
Temperature dependence of singlet-triplet Kondo conductance peaks. (a) Experimental temperature dependence of the conductance of
a diradical molecule (red circles) and its fit by the shown empirical equation (red line)
with resulting values $s=0.7$, $T_\mathrm K=0.6$~K and $G_0=4.9$~nS.
 The temperature dependence strongly deviates from the universal
curve of a spin-1/2 Kondo effect with $s=0.22$ (dashed blue 
line). (b) Theoretical temperature dependence
of the double quantum dot's conductance. Large plot reproduces Figure~\ref{g6} showing the
effect of anisotropic exchange (thin lines) and the corresponding fits (bold symbols) by Eq.~\eqref{eq:universal} (with the exponent $s$ as a free parameter) for restricted temperature ranges resembling the experiment in panel (a). Small plots show the same temperature dependences of renormalized conductance ($\widetilde G_0$ denotes
the fitted conductance span in the given interval) for corresponding values of $\Delta$ (by color coding)
only in the restricted temperature intervals. NRG conductances (points) are fitted (continuous lines) by Eq.~\eqref{eq:universal} with the optimal values of $s$ stated above the plots. Obviously, the increasing values of $\Delta$ give by this fitting procedure larger values of $s$.}
\end{figure*}

The low-temperature behavior of the singlet-triplet Kondo effect is equivalent
with the low-temperature
behavior of a standard spin-half Kondo effect \cite{Pustilnik2000}.
Our theoretical results on the DQD confirm this equivalence, as
long as $T\ll I$ (see Figure~\ref{g2}).
Accordingly, the linear conductance as a function of temperature
should approximately obey the well-known universal curve \cite{goldhaber1998kondo}
\begin{equation}
G(T) = G_{0} \left[1+\left(2^{1/s}-1\right)\left(\frac{T}{T_{0}}\right)^{2} \right]^{-s} + G_{\text{b}},
\label{eq:universal}
\end{equation}
where $T_{0}$ is the approximate 
Kondo temperature, $G_{\text{b}}$ the background conductance, and
$s=0.22$. To experimentally obtain $G(T)$, we recorded d$I$/d$V$ spectra at
fixed $B=6.6$~T and $V_{\text{g}}=-2.8$~V at various temperatures. The linear
conductance was determined by fitting the Kondo peaks to Lorentzian functions
and by extracting the peak height to estimate $G(T)-G_{\text{b}}$. The obtained
values are normalized to $G_{0}$ and plotted in Figure \ref{g8}a, along with the
universal curve with the spin-1/2 value $s=0.22$ (blue dashed line) and a
modified universal curve with $s=0.7$ (red full line). Remarkably, the
experimental data strongly deviates from the universal curve for a standard
spin-1/2 system ($s=0.22$). A good agreement with the experimental data is found
by choosing a significantly higher value for the empirical parameter $s$, which
illustrates the anomalous behavior of this Kondo effect.

Here, we propose an explanation for the anomalous temperature dependence, based
on comparison with the theoretical results from previous sections.
The main panel in Figure~\ref{g8}b 
shows how $G(T)$ (thin solid lines) in NRG calculations is influenced by increasing the
anisotropic exchange $\Delta$. The low-temperature conductance decreases for
higher $\Delta$ and a bump appears at $T \sim 10^{-5}D$ for the blue and green
curves. We find that in a restricted temperature range, the NRG curves can be
well approximated by Eq.~\eqref{eq:universal} with $s>0.22$. The
corresponding fits are drawn in the main panel of Figure~\ref{g8}b as thick solid
lines. The small panels of Figure~\ref{g8}b show the normalized NRG results (plus
signs) for each $\Delta$, along with the fit (solid line) to Eq.~\eqref{eq:universal}
and the corresponding $s$ values. This analysis effectively shows that
nonzero values of $\Delta$ may result in significantly higher values of $s$ coming from the fits.
From this observation we conclude that the presence of an anisotropic exchange
interaction is a possible explanation for the anomalous temperature dependence
of the singlet-triplet Kondo effect observed in this experiment.

\subsubsection{Discussion of the Temperature Dependences}
The larger values of $s$ which result from
fitting the theoretical temperature
dependences [Figure~\ref{g8}b, main panel]
can be attributed to two effects: First,
the $G(T)$ does not reach the maximum value $G_0$ 
due to the anisotropic exchange between the two spins.
Second,
the high-temperature minimum of $G(T)$ in the restricted
temperature range is larger than in the
standard case (\emph{i.e.}, SIAM), because of the
bump caused by spin excitations.
We conclude that the interval of temperatures in which
an apparently anomalous behavior can be observed is set
by two energy scales: $\Delta$ and $I$.
It follows that for the
molecular junction studied here,  
$I \gtrsim 0.4$~meV (corresponding to the highest
temperature 4.2~K). This bound is consistent
with the value of $I=0.7$~meV given by the
zero-field splitting in Figure~\ref{g7}.
Similarly, we can estimate $\Delta$: based on Figure~\ref{g6} we
deduce that $\Delta$ marks the onset of the conductance decrease.
Consequently, from Figure~\ref{g8}a we get $\Delta\approx 0.4\, T_0
= 20\ \mu\text{eV}$.
This value is more difficult to compare.
We remark that spin-orbit interaction in planar graphene-related systems
also lies in the sub-milielectronvolt range \cite{CastroNeto2009}.
In principle, lowering the temperature below $\Delta$ could
lead to splitting of the zero-bias peak as seen in  Figure~\ref{g6},
allowing the more precise determination of $\Delta$.

As we argued in the Methods section, the chosen form of
the AX in Eq.~\eqref{eq:ax} is not generic and other
terms (such as $\hat S_1^x\hat S_2^y$)
can be expected in the molecular junction. Arguments based on
molecular symmetry are not applicable here because the molecular
geometry is in general distorted due to binding to the leads.
Moreover, anisotropic $g$-tensors could also result from spin-orbit
interaction. Naturally, these different anisotropy terms can not be
easily disentangled in a transport measurement. However, as long as the
anisotropies are weak (compared to $I$), their main effect is the avoided
crossing with the energy scale $\Delta$.
On the basis of these considerations we propose that the anomalous
temperature dependence observed in the molecular diradical junction is
caused by spin anisotropy terms with a characteristic
energy scale $\Delta\approx 20\ \mu\text{eV}$ (for
a given direction of the external magnetic field).

As an alternative scenario of the anomalous temperature
dependence of the conductance we mention two-channel Kondo (2CK) physics.
As we argued in Methods section, the 2CK effect can be manifested in a smaller
portion of the parameter space. Mitchell \emph{et al.}\cite{Mitchell2017}
found a temperature dependence similar to that in Figure~\ref{g8} at
the so-called quantum-interference node. The latter represents
a special point of the molecular 2CK, which can be reached
by tuning the gate voltage. In our case, the gate voltage was not
tuned, and therefore, we think that an intrinsic mechanism, the
anisotropic exchange, is more plausible. Moreover, the theoretical
results of this work apply to a wider parametric window.

\section{Conclusions}
We have analyzed the double quantum dot model 
at the singlet-triplet crossing in the regime of
strong quantum fluctuations (Kondo effect) with the
numerical renormalization group. We have focused on the shift of the singlet-triplet degeneracy, which 
can be interpreted as an effective magnetic field generated
by the leads. Its parametric dependence is non-trivial (apparently non-perturbative),
pointing to the role of strong quantum fluctuations.
When the external magnetic field is tuned to the degeneracy
and Kondo plateaus emerge in the conductance, the two dots still exhibit a sizable
spin-polarization. This is surprising, in view of the traditional 
picture of Kondo-screened moments. 

Furthermore, we have studied the effect of an anisotropic exchange (AX). Our data shows
that the singlet-triplet Kondo effect is stable  against weak
AX. The AX of the order of $T_\mathrm K$ causes lowering
of the Kondo plateaus in the temperature dependence of the
conductance, $G(T)$.
The calculated temperature dependences $G(T)$ bear a strong imprint
of the two low-energy scales, $\Delta$ and the exchange $I$.

We have presented experimental data on a molecular junction containing
an organic di-radical coupled to Au leads. The differential conductance
as a function of magnetic field shows a characteristic fingerprint
of the singlet-triplet splitting. The zero-bias resonance at the
singlet-triplet degeneracy point has a temperature dependence which
deviates greatly from the universal curve expected for standard
spin-1/2 Kondo systems. We propose an explanation based
on the lowering of the conductance plateaus caused by anisotropy terms.

\subsection*{Acknowledgments}
We gratefully acknowledge support from the PRIMUS/Sci/09 programme of the
Charles University (R.K.) and from the Czech Science Foundation by Grant
No.~16-19640S (T.N.\ and P.Z.). The experimental work (J.dB., R.G., H.S.J.vdZ.)
was supported by the Netherlands Organisation for
Scientific Research (NWO/OCW), as part of the Frontiers of Nanoscience program,
and the ERC Advanced Grant agreement number 240299 (Mols@Mols).
We thank J.~Veciana and C.~Rovira for the synthesis of the diradical molecule and M.~\v{Z}onda for his substantial help with setting up the NRG calculations at the beginning of the project.


\providecommand{\latin}[1]{#1}
\makeatletter
\providecommand{\doi}
  {\begingroup\let\do\@makeother\dospecials
  \catcode`\{=1 \catcode`\}=2 \doi@aux}
\providecommand{\doi@aux}[1]{\endgroup\texttt{#1}}
\makeatother
\providecommand*\mcitethebibliography{\thebibliography}
\csname @ifundefined\endcsname{endmcitethebibliography}
  {\let\endmcitethebibliography\endthebibliography}{}


\end{document}